\documentclass[%
 reprint,
prb,
floatfix,
]{revtex4-2}

\usepackage{graphicx}
\usepackage{dcolumn}
\usepackage{bm}
\usepackage{natbib}
\usepackage{ulem}
\usepackage{xcolor}
\usepackage{amsmath} 
\usepackage{hyperref}
\usepackage{multirow}

\definecolor{newtext}{RGB}{0, 0, 0}

\begin{document}

\preprint{APS/123-QED}

\title{Unidirectional propagation of spin waves excited by femtosecond laser pulses\\ in a planar waveguide 
}

\author{P. I. Gerevenkov}
\email{petr.gerevenkov@mail.ioffe.ru}
\homepage{http://www.ioffe.ru/ferrolab/}
 \affiliation{Ioffe Institute, 194021 St. Petersburg, Russia}
\author{Ia. A. Filatov}
 \affiliation{Ioffe Institute, 194021 St. Petersburg, Russia}
\author{A. M. Kalashnikova}
 \affiliation{Ioffe Institute, 194021 St. Petersburg, Russia}
\author{N. E. Khokhlov}
 \affiliation{Ioffe Institute, 194021 St. Petersburg, Russia}

\date{\today}

\begin{abstract}
Low-energy magnonic logic circuits are an actively developing field of modern magnetism.
The potential benefits of magnonics for data processing are vitally dependent on units based on nonreciprocal propagation of spin waves, in analogy to semiconductor diodes and transistors in electronics.
In this article, we suggest the approach to realize nonreciprocal propagation of spin waves in a ferromagnetic metallic waveguide by exciting them with femtosecond laser pulse.
Using micromagnetic modeling, we show that the combination of an external magnetic field and the position of the excitation laser spot across the waveguide leads to unidirectional propagation of the excited spin-wave packet.
The results are meaningful for the design of hybrid magnonic-photonic circuits in future generations of data processing devices.
\end{abstract}

\maketitle


\section{\label{sec:Introduction} INTRODUCTION }

Low-energy magnonic logic circuits are an actively developing field of modern magnetism~\cite{chumak2022roadmap,barman20212021,chumak2015magnon,nikitov2015magnonics}.
Short wavelengths of spin waves (SWs) provide scalability of magnon logic units down to tens of nanometers~\cite{manipatruni2018beyond, wang2019spin, wang2020magnonic}.
Furthermore, magnonics paves the way for simplifying logic gates and transferring information using not only the amplitude but also the phase of SWs~\cite{wang2020magnonic, fischer2017experimental, mahmoud2020introduction}.
This is also promising for the neuromorphic and reservoir computing frameworks~\cite{papp2021nanoscale, Watt_ReservoirComputing_PhysRevApplied2020,Watt_Enhancing_JAP2021, Kruglyak_Chiralmagnonic_APL2021, Arai_JAP2018, Arai_PRApplied2018}.
Moreover, the natural nonreciprocity in amplitude and frequency of SWs~\cite{damon_magnetostatic_1961, gallardo2019reconfigurable, gallardo2019spin} promotes the usage of SWs in computing circuits.
For example, the feature is crucial for the realization of unidirectional magnionic circuits, diodes, and isolators~\cite{szulc2020spin, grassi2020slow, shichi2015spin}.
Interest in practical aspects of magnonics has led to the emergence of many proof-of-concept prototypes of nonreciprocal devices over the past few years, such as half-adder~\cite{wang2020magnonic}, majority gate~\cite{talmelli2020reconfigurable, fischer2017experimental}, transistor~\cite{chumak2014magnon}, circulator~\cite{szulc2020spin}, and others.
Such devices are based on the propagation of SWs in waveguides, driven by radio frequency (rf) antennas, most commonly ~\cite{demidov2015magnonic,sadovnikov2015magnonic, khivintsev2022spin}.
However, other options for SW excitation are actively developing: by magnetoelectric transducers~\cite{Cherepov_MEgenerationSW_APL2014, dutta2015non, Duflou_MEforSW_APL2017, vanderveken2020excitation, Mahmoud_Would2022,azovtsev2019dynamical,azovtsev2021energy}, spin-torque oscillators~\cite{demidov2010direct, demidov2016excitation}, and femtosecond laser pulses~\cite{satoh2012directional, Au_directExcitation_PRL2013}, for example.
In the latter case, a wide spectrum of SW frequencies is excited, and the wave propagates in the form of a spin-wave packet~\cite{KamimakiPRB:2017, Filatov_SWinFe_APL2022, filatov2020spectrum, khokhlov2019optical}.
The propagation of such a laser-excited wavepacket even in continuous films is accompanied by a specific spectrum evolution, directionality of propagation, etc., stemming from spatial localization of the laser-induced driving force~\cite{Au_directExcitation_PRL2013, khokhlov2019optical, Jackl_PhysRevX2017, Muralidhar_caustics_PRL2021}.

Laser-induced SW dynamics has just begun to be explored in confined elements for magnonics, such as resonators~\cite{khramova2022accumulation} and in arrays of ferromagnetic nanostripes~\cite{Pal_SWinPy_AdvMatIt2022, Godejohann_MagmonPolaron_PRB2020, Salasyuk_PRB2018}.
However, the case of an individual SW waveguide under optical excitation has not yet been sufficiently explored.
In contrast to infinite films, the contribution of the dipole field in confined structures leads to inhomogeneity of the magnetization distribution near the edges and thus to \textcolor{newtext}{some features} strongly affecting SW characteristics.
For example, edge effects in waveguides reduce SW frequencies with distance from the center, providing selective excitation of central or edge modes of SW with rf antennae~\cite{demidov2015magnonic, Demidov_NanoopticsSW_APL2008, Zhang_SWfrequencyDivision_APL2019}.
One anticipates that localization of the excitation area with a focused laser pulse could give the possibility to excite edge or central SW modes selectively.
Because of the broadband nature of pulsed optical excitation, engineering dispersion of the medium outside the laser spot could enable control of SW propagation.
A waveguide is the simplest example of such engineering, with internal magnetic fields being inhomogeneous across its width. 
Therefore, it is an appealing idea to combine the selectivity of localized optical excitation with profiled magnetic properties of a waveguide to steer generated SWs in it.

In this paper, we consequently examine SW dynamics in the waveguide subjected to broad-area laser pulse excitation, excitation with elongated laser spot resembling rf antennae geometry, and localized laser spot.
We show that the strength of the external magnetic field controls the spatial position of the maximum SW excitation and enables its shift from the center to the edges of the waveguide.
Based on these findings, we demonstrate that a combination of the external magnetic field and position of the laser spot across the waveguide leads to tunable nonreciprocity and even unidirectionality in the propagation of the excited SW wavepacket.
The results are meaningful for the design of hybrid magnonic-photonic circuits in future generations of data-processing devices~\cite{li2020hybrid}.

The paper is organized as follows.
In Sec.~\ref{sec:Model_details} we describe the methods and modeled waveguide parameters.
Section~\ref{subsec:Static_characteristics} presents calculations of the magnitude of magnetization dynamics triggered by laser-induced heating.
In Sections~\ref{subsec:2D_excitation}-\ref{subsec:0D_excitation} we analyze the magnetization dynamics in the waveguide in the case of uniform, elongated, and round-spot ultrafast heating with femtosecond laser pulse sequentially.
In Section~\ref{subsec:coeff_norecipr} we analyze the nonreciprocal propagation of SWs quantitatively and compare our results with the previously reported ones. 
The implementation of the observed tunable nonreciprocity to neuromorphic computing and logic gates is also proposed. 
In the Conclusion we discuss the results obtained, including their possible applications.

\section{\label{sec:Model_details} MODEL DETAILS }

\begin{figure}
\includegraphics[width=0.9 \linewidth]{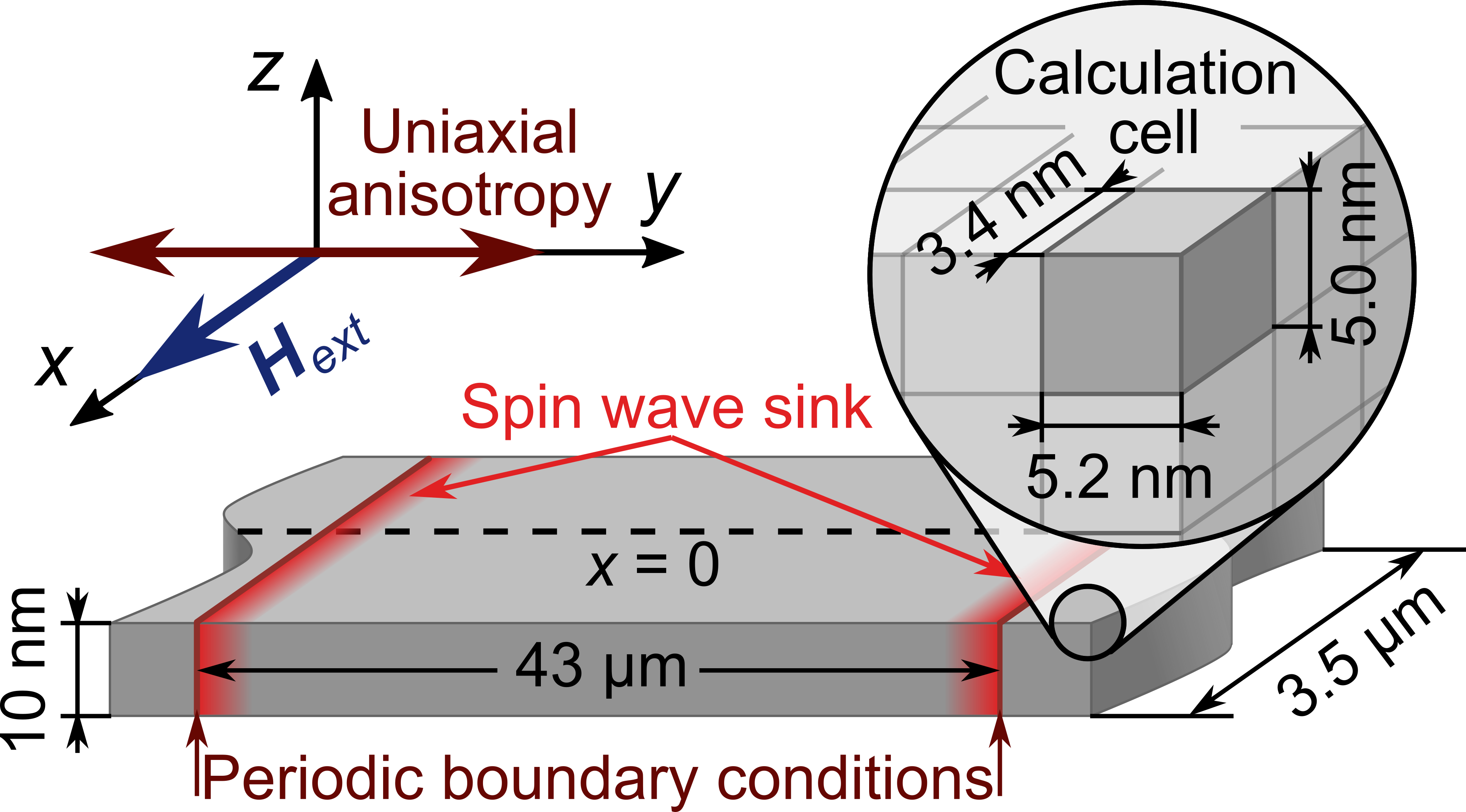}
\caption{\label{fig:model}
Scheme of the simulated waveguide with the easy axis of uniaxial anisotropy along the $y$-axis (shown by the double arrow).
External magnetic field $\mathbf{H}_{ext}$ is applied along the $x$-axis.
The inset shows the cell dimensions used in the simulation.
}
\end{figure} 

We use micromagnetic numerical simulation to model SW propagation in a ferromagnetic waveguide with the width of 3.5\,$\mu$m along the $x$-axis, length of 43\,$\mu$m along the $y$-axis, and thickness of 10\,nm along the $z$-axis (Fig.~\ref{fig:model}).
The chosen waveguide dimensions correspond to those used in the experimentally demonstrated magnonic devices~\cite{talmelli2020reconfigurable,demidov2015magnonic}.
The center of the waveguide is $x = 0$, $y = 0$.
We choose the parameters of the ferromagnetic material corresponding to permalloy at room temperature with saturation magnetization $M_S = 8 \times 10^5$\,A/m, an exchange stiffness parameter $A = 1.3 \times 10^{-11}$\,J/m, and Gilbert damping parameter $\alpha = 0.008$~\cite{zhao2016experimental, Han_DWmotion_bySW_2009_APL}.
There is an additional uniaxial anisotropy with parameter $K_U = 5 \times 10^3$\,J/m$^3$ along the waveguide.
A cell size of $3.4 \times 5.2 \times 5.0$\,nm$^3$  is chosen to be smaller than the magnetostatic $\sqrt{2 A / (\mu_0 M_S^2)} = 5.7$\,nm and magnetocrystalline $\sqrt{A/K_U} = 50$\,nm exchange lengths ~\cite{abo_definition_Lex_IEEE_2013}, where $\mu_0$ is the vacuum permeability.
To exclude edge effects at the $y$-direction ends of the waveguide, the one-dimensional (1D) periodical boundary conditions are used.
Additionally, the value of $\alpha$ exponentially increases to 2.4 within 3\,$\mu$m towards the $y$-edges to avoid back reflection of the spin waves at the ends.
An external magnetic field ${\bf H}_{ext}$ is applied along the $x$-axis with the tilt of $0.1\,^\circ$ both in-plane and out-of-plane of the waveguide to exclude metastable states of magnetization spatial distribution.

We consider the impact of the laser pulse as a relative reduction followed by recovery of the magnetic parameters $K_U$ and $M_S$, as it is observed experimentally in metallic films~\cite{Bigot:PRL1996, Carpene_ultrafast_3D_anisotropy_change_PRB_2010, Gerevenkov_PhysRevMaterials2021}.
It is well established that ultrafast laser-induced heating results in subpicosecond demagnetization in a metal, followed by partial recovery of the magnetization. 
After $\sim2$~ ps, all subsystems of the material are thermalized at an elevated temperature, and the magnetization and anisotropy parameter correspond to this temperature. 
As the period of the precession and SWs excited due to such processes is much larger than this time, it is sufficient to use instantaneous laser-induced reduction of $K_U$ and $M_S$ in the model.
After reduction, the magnetic parameters recover exponentially to their equilibrium values with characteristic time $\tau = 300$\,ps~\cite{Gerevenkov_PhysRevMaterials2021}.
Following experimental observations for metallic films~\cite{Carpene_ultrafast_3D_anisotropy_change_PRB_2010, Gerevenkov_PhysRevMaterials2021, shelukhin2020laser}, the power law $K_U^h/K_U^c = [M_S^h/M_S^c]^a$ is conserved in the entire magnetic parameters' recovery time, where the superscripts $c$ and $h$ correspond to the parameters before and just after excitation, respectively.
Here, we assume the power $a=3$ corresponding to the case of uniaxial anisotropy~\cite{zener1954classical}.
The maximum $M_S$ reduction is assumed to be 10\,\%.
The above process is at least an order of magnitude faster than the period of the excited precession and low-energy spin waves considered in this work.
Therefore, the following set of parameters is used in the model: $M_S^c = 8 \times 10^5$\,A/m, $K_U^c = 5 \times 10^3$\,J/m$^3$ and $M_S^h = 7.2 \times 10^5$\,A/m, $K_U^h = 3.6 \times 10^3$\,J/m$^3$.
The resulting temporal dependence of magnetic parameters is:

\begin{equation}\label{eq:TimeDep}
    P(t) = P^c  - (P^c - P^h) \Theta(t) \mathrm{exp}\left( {\frac{-t}{\tau}} \right),
\end{equation}
where $P$ stands for $M_S$ or $K_U$, and $\Theta(t)$ is a Heaviside function.
In the work, we vary the shape of the excitation laser spot with corresponding changes in spatial distribution $P(x,y)$.
The particular forms of $P(x,y)$ are specified in the corresponding sections below.
Additionally, as demonstrated in experiments of laser-induced thermal change of anisotropy parameters in metallic films, the polarization state of the pump pulse does not affect the excited magnetization dynamics~\cite{khokhlov2019optical, shelukhin2018ultrafast, Gerevenkov_PhysRevMaterials2021}.
Therefore, we do not use the dependence on the polarization of the pump pulses in Eq.~(\ref{eq:TimeDep}).
Also, since the micromagnetic simulation is based on the Landau-Lifshitz-Gilbert equation, the calculations are done effectively at 0\,K.
However, the obtained results would be valid for experiments at room temperature, as the Curie temperature of 10-nm-thick permalloy waveguide is above 600\,K~\cite{Zhang_APLMaterials2019}. 

All calculations are performed using the GPU-accelerated micromagnetic framework mumax$^3$~\cite{vansteenkiste2014design}.
More details of numerical simulations are presented in Sec.~I Suppl. Material~\cite{Supplemental}.

\section{\label{sec:Results_and_discussion} RESULTS AND DISCUSSION }

\subsection{\label{subsec:Static_characteristics} Laser-induced deflection of the effective magnetic field }

\begin{figure}
\includegraphics[width=0.9 \linewidth]{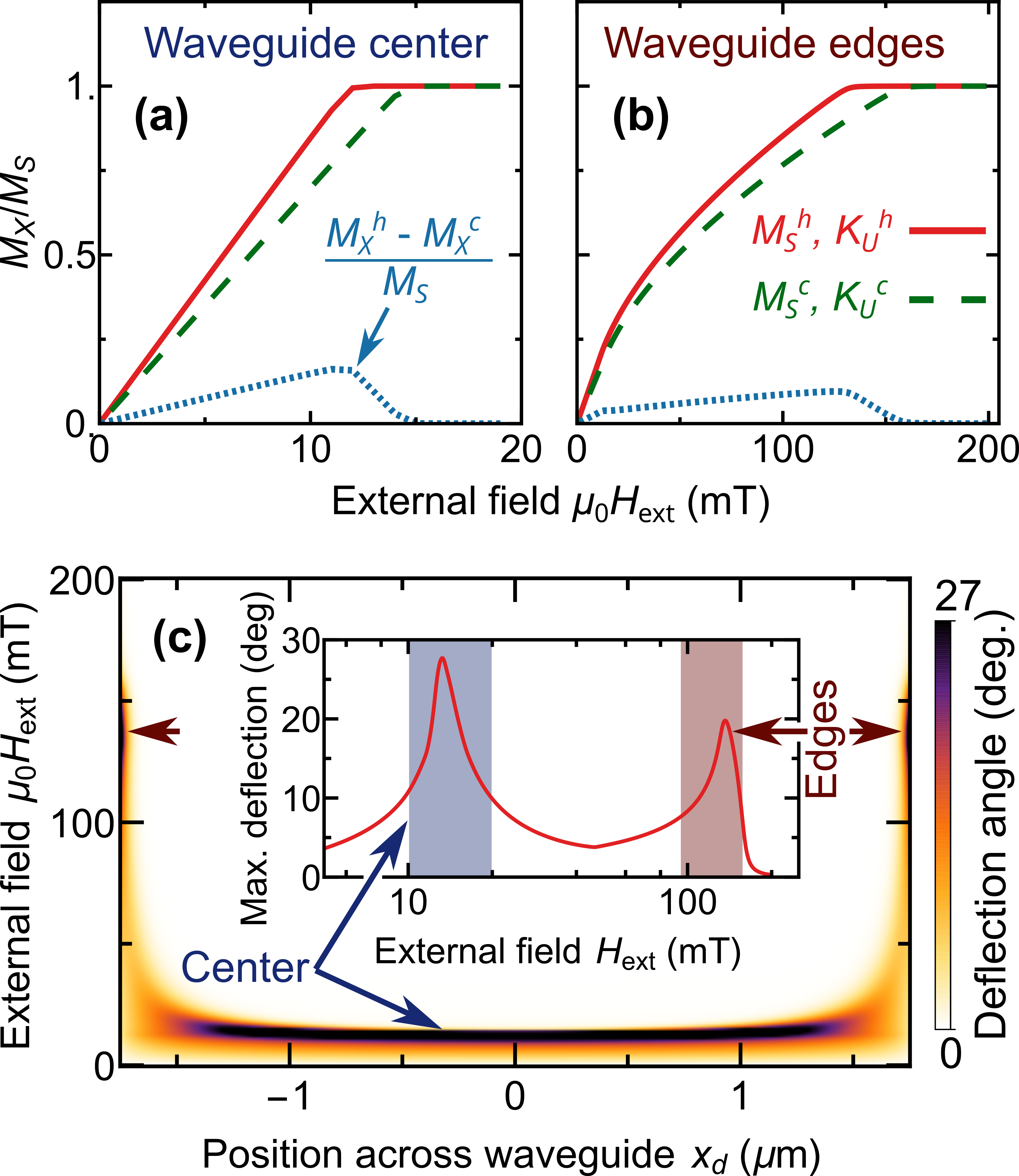}
\caption{\label{fig:Loops}
Deflection of effective magnetic field direction in the case of uniform quasistatic heating.
(a, b) The simulated magnetization curves measured at the center (a) and the edge (b) of the waveguide.
The curves are shown for the parameters of the heated (solid line) and unheated (dashed line) waveguide.
The dotted lines show magnetization deflection in $x$-direction. 
(c) Dependence of deflection angle between magnetization directions for heated and unheated magnetic material as a function of external magnetic field and position across the waveguide.
The inset shows the maximum deflection angle over the whole waveguide width vs. external magnetic field.
}
\end{figure} 

\begin{figure*}
\includegraphics[width= \linewidth]{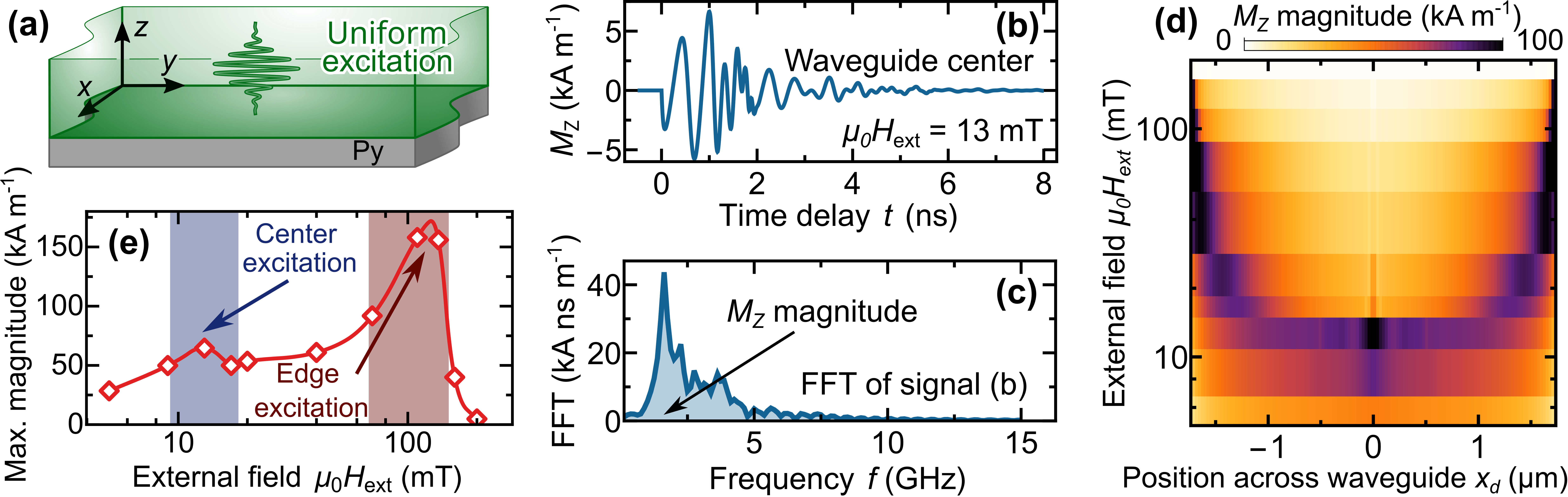}
\caption{\label{fig:2D}
Dynamics of magnetization upon excitation of the entire area of the waveguide.
(a) Waveguide excitation scheme.
(b) An example of a laser-induced magnetization dynamics time-signal at the center of the waveguide at $\mu_0H_{ext} = 13$\,mT.
(c) Fast Fourier transform of time-signal (b).
Estimated signal magnitude is shown by the shaded area.
(d) The signal magnitudes of laser-induced magnetization dynamics as a function of the external magnetic field and position across the waveguide.
(e) The maximum magnitudes of magnetization dynamics over the whole waveguide width versus the external magnetic field.
Symbols are simulation results, the line is the guide for an eye.
}
\end{figure*}

First, we perform the simulation of the laser-induced initial deflection of the total effective field $\mathbf{H}_{eff}$ across the waveguide.
This deflection defines the initial magnitude of the excited magnetization dynamics~\cite{Carpene_ultrafast_3D_anisotropy_change_PRB_2010, khokhlov2019optical,Gerevenkov_PhysRevMaterials2021}. 
The deflection of $\mathbf{H}_{eff}$ is found as the angle between static magnetization in the cold and heated film at various positions $x_d$ across the waveguide and values of ${H_{ext}}$.

Fig.~\ref{fig:Loops}\,(a,b) show magnetizing curves probed at the center and $x$-edges of the waveguide for two sets of parameters: $M_S^c, K_U^c$ (dashed line) and $M_S^h, K_U^h$ (solid line).
The maximum deflection of the magnetization vector in the center is observed when $\mu_0 H_{ext}$ is close to the anisotropy field $2K_U/M_S$ with the demagnetizing field addition (Fig.~\ref{fig:Loops}\,(a)).
The demagnetizing field and consequently the external field corresponding to the maximum magnetization deflection increase by an order of magnitude closer to the $x$-edges of the waveguide~(Fig.~\ref{fig:Loops}\,(b)).

Fig.~\ref{fig:Loops}\,(c) shows the angle of deflection between $\mathbf{M}^c$ and $\mathbf{M}^h$ (or deflection of $\mathbf{H}_{eff}$) as a function of ${H}_{ext}$ and the position of the detection region across the waveguide (the color map chosen from~\cite{kovesi2015good} is darker with a higher deflection).
At a distance of the order of 100\,nm from the $x$-edge, an abrupt shift of the maximum deflection to larger values of ${H}_{ext}$ is observed.
Thus, two peaks of the maximum deflection angle among all detection positions vs. ${H}_{ext}$ present (insert in Fig.~\ref{fig:Loops}\,(c)).
The dependence is obtained from the map in Fig.~\ref{fig:Loops}\,(c) by choosing the maximum of the deflection angle at each ${H}_{ext}$ value.
The first peak at 13\,mT corresponds to the center region of the waveguide, and the second one at 136\,mT corresponds to the edge area.

Above, we considered quasistatic heating.
It means, the magnetization vector follows the $\mathbf{H}_{eff}$.
In the case of ultrafast heating by the laser pulse, the direction of $\mathbf{H}_{eff}$ changes much faster than the period of the precession, and the magnetization becomes noncollinear with the new orientation of the effective field~\cite{Kirilyuk_RevModPhys2010}.
Because of that, magnetization starts to precess around a new orientation of $\mathbf{H}_{eff}$ due to the torque $\mathbf{T}$ acting on $\mathbf{M}$ following the Landau-Lifshitz equation $\mathbf{T} \sim \left[ \mathbf{M} \times \mathbf{H}_{eff} \right]$~\cite{landau1935theory}.
Thus, the magnitude of $T$ is proportional to the sine of the deflection angle, obtained above.
Therefore, since $\mathbf{H}_{eff}$ has the local contribution of demagnetizing fields varying to the edges, the value of $H_{ext}$ defines the localization of SW excitation upon ultrafast laser-induced heating, following Fig.~\ref{fig:Loops}\,(c).

Further, we are motivated by the experiments with rf antennas and single-frequency excitation, which show the selectivity of SW excitation regions~\cite{Demidov_NanoopticsSW_APL2008, demidov2015magnonic, Zhang_SWfrequencyDivision_APL2019}.  
Since a femtosecond laser pulse excites a wide frequency spectrum of magnetization dynamics but in a local area, the following question arises: how do laser-induced wide-spectrum  SW packets propagate after optical excitation in a confined waveguide, and could we control whether the central or edge mode is excited by positioning the excitation area and value of ${H}_{ext}$? 
Below, we examine the SW propagation at different shapes and positions of the excitation area and various values of ${H}_{ext}$.

\subsection{\label{subsec:2D_excitation} Magnetization dynamics at uniform laser excitation} 

First, we consider the uniform laser excitation of the whole waveguide area~(Fig.~\ref{fig:2D}\,(a)).
To detect excited precession with a resolution across the waveguide, we average magnetization dynamics over the entire simulated length and 42\,nm width regions at various positions $x_d$~(see details in Sec.~II in Suppl. Material~\cite{Supplemental}).
The temporal dynamics, probed at different $x_d$ across the waveguide, is rather complex and exhibits a wide frequency spectrum (Fig.~\ref{fig:2D}\,(b,c)).
It is because the edges of the stripe act as sources of nonzero value of torque $T$ exciting SWs with different frequencies even at uniform external stimuli (see Ref.~\cite{au2011excitation, Davies_generation_IEEE2016, Trager_kselectiveSW_Nanoscale2020, Trager_CompetingSWemission_PRB2021} and Sec.~IV in Suppl. Material~\cite{Supplemental}).
To characterize excitation efficiency, we introduce, following Refs.~\cite{khokhlov2021neel, Gerevenkov_PhysRevMaterials2021}, the magnetization dynamics integral magnitude as the area under the FFT curve (shaded area in Fig.~\ref{fig:2D}\,(c)).
It should be noted that the obtained magnitude includes not only uniform precession, but also SWs from the edges and higher waveguide modes.
Consistent with calculations presented in the previous section, the maxima of magnetization dynamics magnitudes are found at the waveguide center at low $H_{ext}$ and at the $x$-edges at higher values of $H_{ext}$~(Fig.~\ref{fig:2D}\,(d)).

The dependence of the maximum precession magnitude versus external field shows two peaks at 13 and 136\,mT~(Fig.~\ref{fig:2D}\,(e)), where the maximal laser-induced deflection of $\mathbf{H}_{eff}$ is obtained at the center and at the edges (insert in Fig.~\ref{fig:Loops}\,(c)).
One can see that the maximum magnitude in the center of the waveguide appears to be lower than the one at the edges. 
This is in contrast to the maximum deflection angles at these positions (Fig.~\ref{fig:Loops}\,(c)). 
Since we use the out-of-plane component of magnetization as a measure of the precession magnitude, this discrepancy is partly due to the different ellipticities of precession across the waveguide.
Another reason is the different ratio between the precession period and $\tau$.
At larger $H_{ext}$ the precession period is less than $\tau$ and the initial amplitude is proportional to the laser-induced effective field deflection angle~\cite{shelukhin2018ultrafast,Gerevenkov_PhysRevMaterials2021}.
At lower $H_{ext}$ the precession period is larger than $\tau$ and the initial amplitude becomes a function of both the deflection angle and precession frequency~(see details in Sec.~IV of Suppl. Material~\cite{Supplemental} and~\cite{li2022perspective}).

The obtained results demonstrate the possibility of laser-induced excitation of different areas across the waveguide depending on the value of ${H}_{ext}$.
It is in agreement with the torque $\mathbf{T}$ spatial distribution predicted in the previous Section.
Furthermore, the selectivity of the excitation region could be realized with the pump laser positioning across the waveguide and the spot’s shape.
Hereafter, we show that such an approach enables one to control the nonreciprocity of SW propagation and even to achieve its unidirectionality.

\begin{figure*}
\includegraphics[width= 0.9 \linewidth]{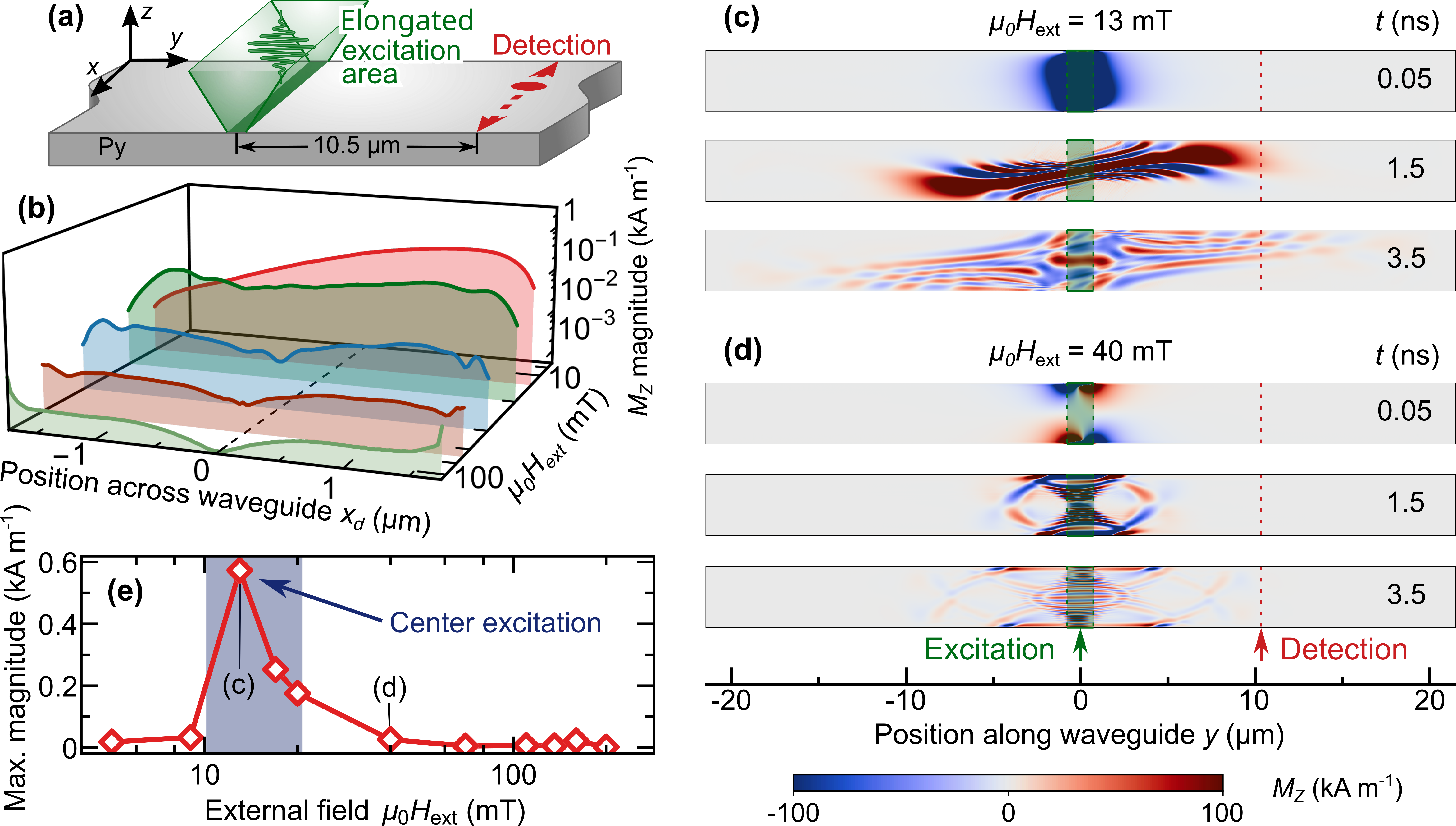}
\caption{\label{fig:1D}
Propagation of SWs upon elongated spot excitation.
(a) Waveguide excitation and detection scheme.
(b) SW magnitude dependences on the detection position across the waveguide for various values of the external field.
(c, d) Distributions of $M_z$ magnetization component at various times $t$ after the excitation for $\mu_0H_{ext} = 13$ and 40\,mT, respectively. 
(e) The maximum magnitudes of excited SWs over the whole waveguide width versus the external magnetic field.
}
\end{figure*} 

\subsection{\label{subsec:1D_excitation} Spin waves' propagation at elongated spot excitation}

Next, we consider the focused excitation spot elongated across the whole waveguide~(Fig.~\ref{fig:1D}\,(a)), which can be achieved using cylindrical lenses or a metallic slit, for example~\cite{Kainuma_Fastacquisition_APExp2021, matsumoto_evanescentSW_PRB2020, Hioki_Bireflection_CommPhys2020, Hioki_coherent_CommPhys2022}.
The excitation spot is uniform along the $x$-axis and is located at $y = 0$.
Thus, the temporal and spatial laser-induced evolution of magnetic parameters has the following form:
\begin{equation}\label{eq:1DGauss}
    P(y,t) = P(t)\exp\left(-\frac{y^2}{2\sigma^2} \right),
\end{equation}
where $P(t)$ has the form (\ref{eq:TimeDep}), and $\sigma = 640$\,nm.
To detect propagating wave packets, we measure magnetization dynamics averaged over a 42 by 27\,nm region at various positions across the waveguide $x_d$.
The detection area is placed at a distance $y_d = 10.5$\,$\mu$m from $y = 0$ in the positive direction of the $y$-axis.
At this distance, the SW dynamics is experimentally detectable using a scanning optical pump-probe technique in the case of unbounded metallic films~\cite{IihamaPRB:2016, kamimaki2017micro, KamimakiPRB:2017, khokhlov2019optical,filatov2020spectrum, Filatov_SWinFe_APL2022} and in permalloy waveguides upon rf field excitation combined with Brillouin light scattering detection~\cite{demidov2015magnonic}.

Fig.~\ref{fig:1D}\,(b) shows SWs packet magnitude versus $x_d$ at different $H_{ext}$.
Following the case of uniform excitation, the dependence demonstrates a single-peak distribution at low $H_{ext}$, but its maximum is shifted from $x_d=0$.
With an increase of $H_{ext}$, the magnitude decreases and its distribution becomes more symmetrical with maxima closer to the waveguide's edges.

\begin{figure*}
\includegraphics[width= 1 \linewidth]{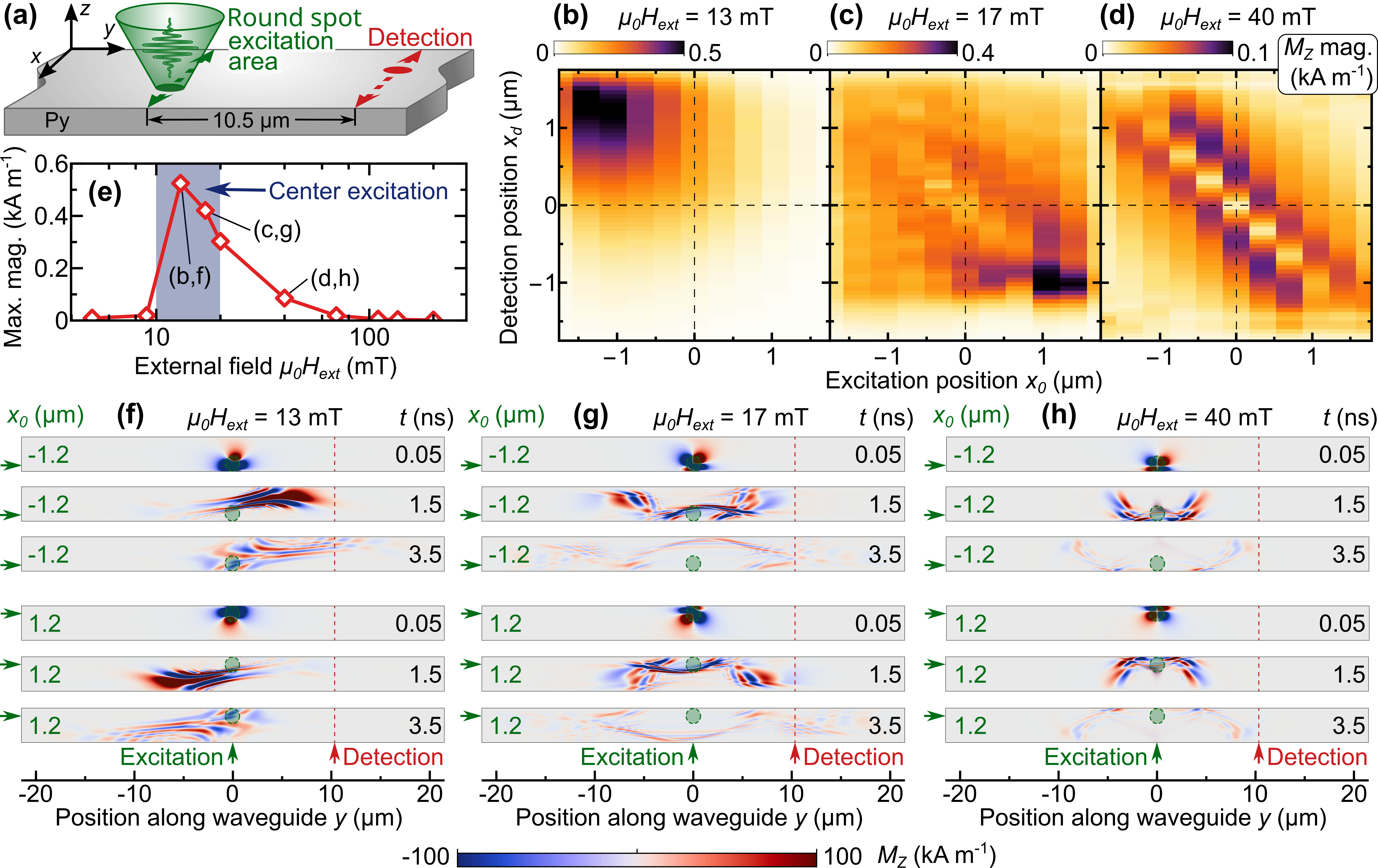}
\caption{\label{fig:0D_1}
Propagation of SWs upon round spot excitation.
(a) Waveguide excitation scheme.
(b - d) The wave magnitude dependences at $y_d = 10.5$\,$\mu$m on the excitation and detection positions across the waveguide for the $\mu_0 H_{ext} = 13, 17$ and 40\,mT, respectively.
(e) The maximum of SW magnitudes over all excitation and detection positions vs. the external magnetic field.
(f - h) Distributions of $M_z$ magnetization component at various times after excitation for $\mu_0 H_{ext} = 13, 17$ and 40\,mT, respectively.
Distributions in the top three rows correspond to excitation at $x_0 = -1.2$\,$\mu$m relative to the waveguide center, and the bottom three are at $x_0 = 1.2$\,$\mu$m.
}
\end{figure*} 

To shed light on the observed features, we analyze the distributions of $M_z$ during the propagation of SWs at low and high $H_{ext}$ (Fig.~\ref{fig:1D}\,(c,d)).
At $\mu_0 H_{ext} = 13$\,mT, the waves propagate from the center at an angle relative to the waveguide long axis (Fig.~\ref{fig:1D}\,(c)).
It is due to the incomplete alignment of the vector $\mathbf{M}^c$ to $x$-axis outside the excitation area at the given external field strength (see Fig.~\ref{fig:Loops} and Sec.~III of Suppl. Material~\cite{Supplemental}).
Since optically excited SWs propagate almost perpendicular to magnetization in thin metal films in the Damon-Eshbach geometry~\cite{damon_magnetostatic_1961, IihamaPRB:2016, khokhlov2019optical}, the propagation direction of the wave packet is deflected from the long axis of the waveguide.
This explains the asymmetry in the magnitude distribution of the SW packets across the waveguide at $y_d$ and at low $H_{ext}$.
Notably, the situation is not specific to optical excitation and is analogous to rf antennae excitation of microwaveguides~\cite{sekiguchi2017spin, vanderveken2020excitation}.

If $H_{ext}$ increases, the efficient excitation area (the area of larger $\mathbf{H_{eff}}$ deflection) shifts to the waveguide edges (Fig.~\ref{fig:1D}\,(d)), where ${\bf M}$ is directed closer to $y$-direction due to demagnetizing fields~(see Sec.~IV of Suppl. Material~\cite{Supplemental}).
Due to this, excited SWs propagate from the edges predominantly along $x$ direction and are detected at $y_d = 10.5$\,$\mu$m after multiple reflections accompanied by the magnitude suppression. 
As a result, the high field peak in the maximum magnitude distribution is suppressed at $y_d$ (Fig.~\ref{fig:1D}\,(e)).

Thus, even if the excitation of the waveguide is homogeneously distributed across its width, sweeping $H_{ext}$ gives the opportunity to tune not only the frequency of SWs, but also their propagation direction in the planar waveguide, and even to suppress them.
However, with the laser pulses as stimuli, we obtain an additional degree of freedom by changing the shape and position of the excitation area. 
The simplest case is to focus the femtosecond laser pulses into a round shape spot using a micro-objective lens.
In the next section, we examine the effect of the position of the round-spot excitation area across the waveguide on the results previously obtained.

\subsection{\label{subsec:0D_excitation} Spin waves propagation at round-spot excitation}

Finally, we consider the excitation by femtosecond laser pulses focused into a 2D-Gaussian spot:

\begin{equation}\label{eq:2DGauss}
    P(x,y,t) = P(t)\exp\left[-\frac{(x-x_0)^2+y^2}{2\sigma^2} \right],
\end{equation}
with $\sigma = 640$\,nm and various positions of the excitation spot center $x_0$ across the waveguide (Fig.~\ref{fig:0D_1}\,(a)).
The detection area is placed at $y_d = 10.5$\,$\mu$m and various $x_d$.

In Fig.~\ref{fig:0D_1}\,(b-d) we plot the SW magnitudes in the detection area as a function of $x_0$ and $x_d$. 
There are high SW magnitudes only at half of the waveguide width at $\mu_0 H_{ext} = 13$ and 17\,mT, corresponding to the efficient excitation of the magnetization dynamics in the center of the waveguide: at $x_d > 0$ if $x_0 < 0$ and 13\,mT (Fig.~\ref{fig:0D_1}\,(b)), and at $x_d < 0$ if $x_0 > 0$ and 17\,mT (Fig.~\ref{fig:0D_1}\,(c)).
If $H_{ext}$ increases, wave packets with similar magnitudes are registered at any $x_0$ across the entire width of the waveguide~(Fig.~\ref{fig:0D_1}\,(d)).
However, in this case, the magnitude rapidly decreases, as illustrated in Fig.~\ref{fig:0D_1}\,(e), where we plot the field dependence of the largest magnitude among the spin waves excited at any $x_0$ and detected at any $x_d$.

Distributions of the $M_z$-component at $\mu_0 H_{ext} = 13$ and 17\,mT~(Fig.~\ref{fig:0D_1}\,(f,g)) demonstrate the propagation of SWs in the positive or negative directions of the $y$-axis depending on the position of the spot $x_0$.
Similarly to the case of elongated excitation (Sec.~\ref{subsec:1D_excitation}), SWs propagate at an angle relative to the waveguide long axis.
For $\mu_0 H_{ext} = 13$\,mT and $x_0 < 0$ the wave packet propagates in the positive direction only of the $y$-axis and vice versa for $x_0 > 0$, i.e. with pure unidirectional propagation of the waves~(Fig.~\ref{fig:0D_1}\,(f)).
The detailed analysis of the origin of the unidirectionality is presented in the next Section.

At $\mu_0 H_{ext} = 17$\,mT, spin waves are excited from the center and from the edges of the waveguide.
It is a result of two mechanisms.
On the one hand, the deviation of ${\bf M}$ from ${\bf H_{ext}}$ is still large enough for the excitation of the center to be nonzero.
But on the other hand, the effective excitation region shifts to the waveguide edge (see Fig.~\ref{fig:2D}\,(d)), which produce edge modes. 
These two mechanisms lead to interference of modes excited from the edge and from the center (see Movie1 file in~\cite{Supplemental}).
Due to interference, the direction of the largest magnitude propagation is reversed with respect to the case at $\mu_0 H_{ext} = 13$\,mT (magnitude's peak at $x_0>0, x_d<0$ in Fig.~\ref{fig:0D_1}\,(c)).

At $\mu_0 H_{ext} = 40$\,mT (Fig.~\ref{fig:0D_1}\,(h)), there are only edge-excited wave packets propagating almost symmetrically in the $y < 0$ and $y > 0$ directions.
It is due to multiple reflections of SWs from the edges of the waveguide, which leads to a decrease in the propagation velocity and detected magnitudes, similar to the elongated excitation case (Sec.~\ref{subsec:1D_excitation}).

\begin{figure*}
\includegraphics[width=0.9 \linewidth]{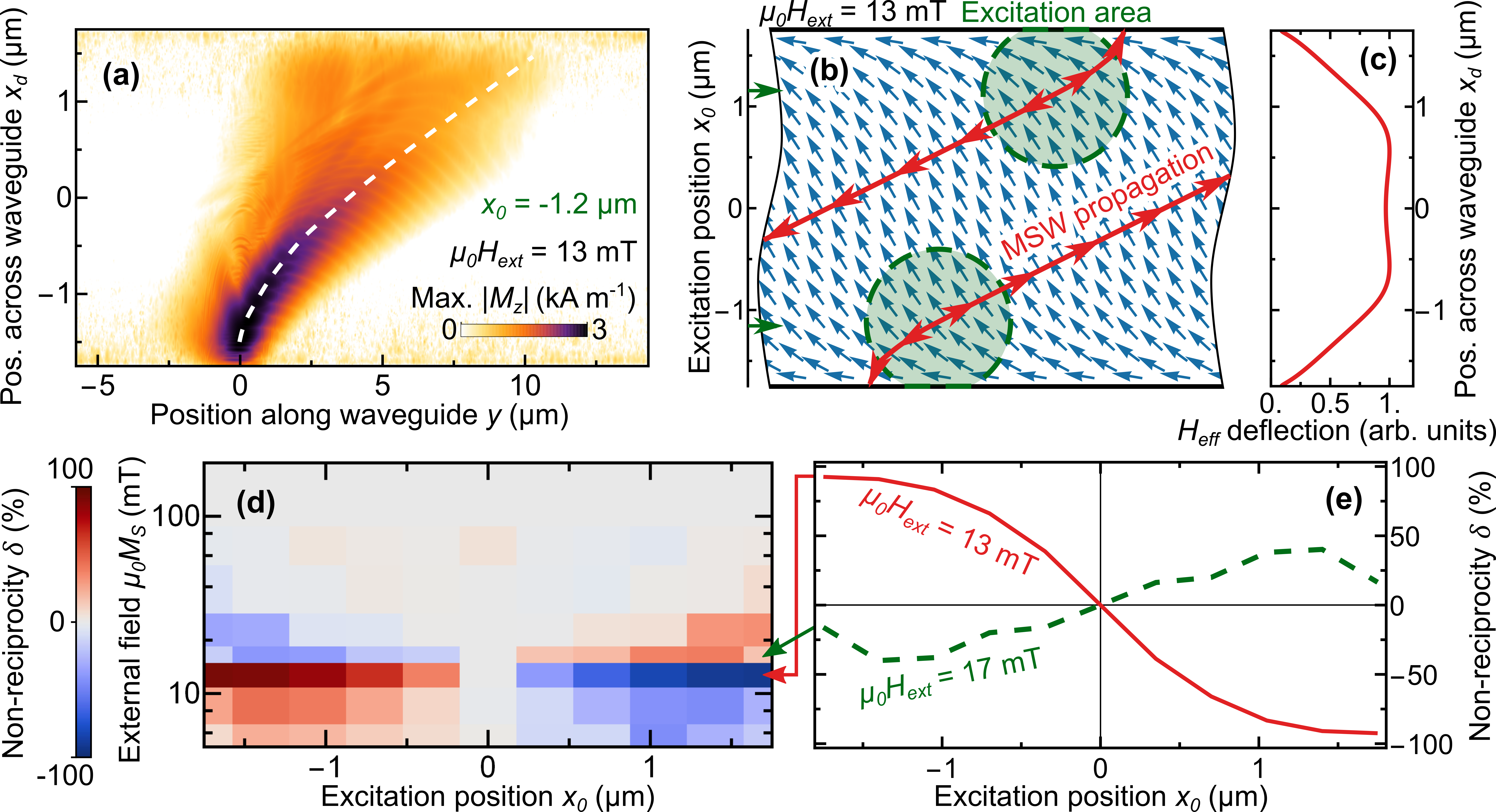}
\caption{\label{fig:Excitation_mechanism}
Excitation mechanism and magnitude nonreciprocity of laser-induced magnetostatic waves.
(a) Distribution of the maximum $|M_z|$ value for the entire simulation time of 8\,ns, on a logarithmic scale.
Data are given for $x_0 = -1.2$\,$\mu$m at $\mu_0 H_{ext} = 13$\,mT.
(b) Small arrows depict magnetization in-plane distribution in the waveguide at $\mu_0 H_{ext} = 13$\,mT.
Shaded areas show the excitation spot FWHM for two $x_0$ positions of $\pm 1.2$\,$\mu$m.
Red arrows show the dominant propagation direction of SWs determined by the normal to the ${\bf M}$ at each point.
(c) Heat-induced change in effective field deflection across the waveguide at $\mu_0 H_{ext} = 13$\,mT.
(d) Nonreciprocity parameter $\delta$ as a function of excitation spot position $x_0$ and $\mu_0 H_{ext}$.
(e) Dependence $\delta(x_0)$ at $\mu_0 H_{ext} = 13$ and 17\,mT.
}
\end{figure*} 

\subsection{\label{subsec:coeff_norecipr} Analysis of unidirectional propagation of spin waves}

Thus, focused femtosecond laser pulses make it possible to excite SWs with unidirectional propagation in confined structures.
In this section, we go deeper into the origins and quantitative analyses of the revealed unidirectionality of the waves.

The most prominent asymmetry in the propagation of the waves is obtained at $\mu_0 H_{ext} = 13$\,mT.
Fig.~\ref{fig:Excitation_mechanism}\,(a) shows the distribution of the maximum $|M_z|$ over the entire simulation time of 8\,ns.
As can be seen, the magnitude of SWs propagating in the positive direction of the $y$ axis is higher than in the negative one.
Hereafter, we argue that unidirectionality is the result of demagnetization fields and symmetry breaking as the excitation area approaches the edge of the waveguide.

To describe the origin of the unidirectionality in detail, we again consider the equilibrium distribution of the magnetization and its initial deflection.
Fig.~\ref{fig:Excitation_mechanism}\,(b) shows the equilibrium magnetization distribution in the waveguide at $\mu_0 H_{ext} = 13$\,mT.
It defines the SW trajectories as those propagating to the waveguide's center in one direction and approaching the edge almost normally in the other direction.
Moreover, as demonstrated in Fig.~\ref{fig:Excitation_mechanism}\,(c), the $\mathbf{H_{eff}}$ deflection angle decreases toward the edges, resulting in efficient excitation of only the central area of the waveguide.
To propagate in the negative $y$-direction, the wave excited in its central part must reach the edge and be reflected from it.
Wave reflection leads to a decrease in magnitude due to nonuniform dispersion across the waveguide and to a change in the propagation direction to the one perpendicular to the $y$-axis.
As a result, edge propagating SWs have almost zero magnitudes at $y_d$.
The mechanism of asymmetric excitation and propagation of SWs is confirmed by the fully symmetric propagation of SWs at $x_0 = 0$ and $\mu_0 H_{ext} = 13$\,mT (see Sec.\,V in Suppl. materials~\cite{Supplemental}).

To provide the quantitative analysis, we introduce a dimensionless nonreciprocity parameter $\delta$ as follows:
\begin{equation}\label{eq:Nonrecip}
    \delta = \dfrac{a_{max}^{+} - a_{max}^{-}}{a_{max}^{+} + a_{max}^{-}},
\end{equation}
where $a_{max}$ is maximum $M_z$ magnitude over all positions of $x_d$ at $|y_d| = 10.5$\,$\mu$m.
The superscript indicates the direction along the $y$-axis in which detection is performed.
Fig.~\ref{fig:Excitation_mechanism}\,(d) shows the parameter $\delta$ as a function of the excitation spot position $x_0$ and $H_{ext}$.
The nonreciprocity increases with an increase in the external field and reaches a maximum at $\mu_0 H_{ext} = 13$\,mT, then changes sign in a field 17\,mT and decreases to zero in fields above 60\,mT.
The function $\delta(x_0)$ is antisymmetric, which means that the SWs change the propagation direction depending on which edge the pump spot is closer to.
Fig.~\ref{fig:Excitation_mechanism}\,(e) shows the cross-sections from Fig.~\ref{fig:Excitation_mechanism}\,(d) at $\mu_0 H_{ext} = 13$ (red solid line) and 17\,mT (dashed green line).
At the field of 13\,mT, the nonreciprocity value is equal to zero when the center is excited and increases up to 93\,\% as the excitation region approaches the edges of the waveguide.
The interference process at $\mu_0 H_{ext} = 17$\,mT results in a change in the sign of $\delta(x_0)$ and suppression of its maximum value at 40\,\% with a slight decrease in the detected magnitude.

The obtained value of nonreciprocity $\delta$ larger than 90\,\% could be regarded as unidirectional propagation.
The recent review on unidirectional spin wave propagation and devices~\cite{chen2021unidirectional_review} gives a broad overview of the phenomena.
Hereafter, we briefly describe the main origins of nonreciprocity to compare our results with the previous one.
We note that our approach is based on symmetry breaking by the excitation technique, not by waveguide properties.
Traditional excitation by rf antennae is also possible to provide the nonreciprocity~\cite{Shibata_APL2018} with the $\delta$ of 50-60\,\% in terms of the Eq.(\ref{eq:Nonrecip}). 
Nevertheless, the asymmetry of the waveguiding structure itself gives rise to the nonreciprocity of SWs.
For example, the nonreciprocity for surface magnetostatic waves is based on the surface properties of the waveguiding medium~\cite{damon_magnetostatic_1961, gurevich1996magnetization}.
This prediction finds the experimental confirmations for different natures of surface asymmetry.
Single waveguide on a substrate reveals the nonreciprocity up to 50\,\% in metallic waveguids~\cite{nakayama2015thickness, jamali2013spin_SciRep, Wong_UnidirectionalAPL2014, Zhang_APL2020} and higher than 90\,\% in dielectric ones~\cite{Wong_UnidirectionalAPL2014, Wang_PRL2020}.
The surface properties could be changed via surface charges at the semiconductor interface~\cite{Sadovnikov_semiconductor_magnonics_PRB2018}, Dzyaloshinskii-Moriya interaction~\cite{Wang_PRL2020}, Seebek effect~\cite{Wang_Seebek_PRL2018}, etc.
The magnetic bilayer structures reveal higher values of magnitude nonreciprocity $\delta$, up to 100\,\%~\cite{shichi2015spin, gallardo2019reconfigurable}.
Periodic magnonic structures, named magnonic crystals, provide nonreciprocity up to 30\,\%~\cite{hara2022intensity_JPD}, but in combination with the bilayer concept the parameter $\delta$ reaches 100,\%~\cite{Chen_PRB2017}.
The interaction of spin waves with the acoustic ones gives near 100\,\% of amplitude nonreciprocity of the coupled magnetoelastic waves~\cite{Tateno_PRApplied2020, Kuss_PRApplied2021, Piyush_SciAdv2020}.

The distinctive feature of the optical approach to SW excitation is its tunability on demand.
Furthermore, the shape of the obtained function $\delta(x_0)$ is smooth and nonlinear (Fig.~\ref{fig:Excitation_mechanism}\,(e)), similar to the sigmoid function.
The last one is the most commonly known function used in artificial neural networks~\cite{han1995influence}.
Thus, the obtained here function $\delta(x_0)$ could be utilized in the design of magnonic-photonic neural networks.
Demonstration of the workflow of such a network could be the next step of research, similar to studies of spin-wave-coupled spin-torque oscillators~\cite{Arai_JAP2018, Arai_PRApplied2018}.
On the other hand, the demonstrated tunability of SW propagation could be used to change the interference pattern of SWs, excited with two and more sources.
This paves the way for the construction of magnonic logic gates for binary, like XOR, NOR, etc, and even nonbinary data processing~\cite{chumak2015magnon, mahmoud2020introduction}.

\section{\label{sec:Conclusion} CONCLUSION}

We demonstrate a number of features of SW propagation at optical excitation in the ferromagnetic metallic waveguide.
From the simulation results of the optical excitation of the entire waveguide area (Sec.~\ref{subsec:2D_excitation}) we obtain two pronounced excitation peaks of spin wave dynamics at external fields of 13\,mT, and 136\,mT.
They correspond to the excitation of the dynamics in the central area (13\,mT) and edges (136\,mT) of the waveguide.
In the case of excitation of the entire waveguide area, the SWs propagate from the edges of the waveguide, as in previous experiments with rf antennae~\cite{Trager_DirectImaging_PSSRRL2020, au2011excitation}.
It is in contrast to the optical excitation of an unbounded film, where propagating SWs are observed only at excitation with tightly focused femtosecond laser pulses~\cite{satoh2012directional, Au_directExcitation_PRL2013, Jackl_PhysRevX2017}. 
Here, the edges of the waveguide serve as sources of stray field, the change of which gives rise to the torque $\mathbf{T}$ acting on the magnetization upon ultrafast heating.

The excitation with a focused laser spot elongated across the whole waveguide also demonstrates center- or edge-excitation of SW packets, depending on the external magnetic field magnitude.
At low field, the packet from central excitation propagates along the waveguide at some angle to its long axis because of incomplete magnetization alignment with the external magnetic field outside the excitation area (Fig.~\ref{fig:Excitation_mechanism}, b).
If necessary, the tilt could be compensated using ferromagnetic dots located near the waveguide, as demonstrated in Ref.~\cite{zhang2019controlled}.
Excitation in higher fields produces SWs propagating mostly perpendicular to the long waveguide axis.
In this case, waves repeatedly reflect from the edges of the waveguide and rapidly decay with increasing $y_d$.
Thus, the external field modulation could be used for suppressing the SW propagation in a waveguide.

An excitation with pulses focused into a round spot smaller than the waveguide width brings an additional degree of freedom to manipulate the parameters of the SWs -- the position $x_0$ of the pump spot across the waveguide.
In addition to the effects observed at elongated spot excitation, nonreciprocal and even unidirectional wave propagation is observed if the pump area approaches the waveguide edge.
We explain these effects by breaking the symmetry of propagation conditions for the SWs excited near the edge of the magnetic medium when the wave propagates toward the center in one direction and towards the edge of the waveguide in the opposite direction.
At low field supporting excitation of SWs in the central part of the waveguide, we achieve nonreciprocity exceeding 90\%, i.e. unidirectional propagation.
At field values that support simultaneous excitation of the SWs from the center and from the edges of the waveguide, the interference between them leads to the change in the nonreciprocity sign, a twofold decrease of its value, and a slight decrease in SW amplitude.
Thus, in the waveguide with nonuniform internal fields across its width, a combination of external field and spot position make it possible to tune the SW propagation from symmetrical to unidirectional and even suppress it.
We believe that the experimental demonstration of the proposed unidirectionality of SWs is quite straightforward in the metallic waveguides with the all-optical pump-probe technique~\cite{Au_directExcitation_PRL2013, khokhlov2019optical, KamimakiPRB:2017, yun2015simultaneous, Filatov_SWinFe_APL2022}.
The waveguide should be wider than the pump spot, but not too wide to have sufficient demagnetizing fields across the waveguide.
Thus, a width of 2--10~$\mu$m, which is commonly used in magnonic element prototypes would be suitable in the case of a metal waveguide.
The measured value would be the time signal of the Kerr effect with further processing to the SW amplitudes and parameter $\delta$, as discussed above.
Although we present the results of numerical calculation here, the optically excited SWs are readily detectable at distances greater than 10 $\mu$m with the pump-probe technique~\cite{khokhlov2019optical,filatov2020spectrum, Filatov_SWinFe_APL2022, IihamaPRB:2016, kamimaki2017micro, KamimakiPRB:2017} or Brillouin light scattering~\cite{demidov2015magnonic, Muralidhar_caustics_PRL2021}.

The effects demonstrated here \textcolor{newtext}{open up ways} of creating magnonic logic elements~\cite{mahmoud2020introduction}.
In particular, local detection of SWs with rf antennae at the waveguide edge has already been proposed for concepts of logic gates performing different Boolean functions~\cite{sekiguchi2017spin}.
Optical excitation could improve the gates' functionality with reconfiguration on demand, which is vital for neuromorfic circuits.
Furthermore, optical impact is not limited to demagnetization or magnetic anisotropy reduction discussed here, and includes various optomagnetic and photomagnetic effects, ultrafast phase transitions, generation of coherent phonons, etc. depending on the materials and experimental conditions~\cite{Kirilyuk_RevModPhys2010, kimel2020fundamentals, bossini2016magnetoplasmonics, Jager_Picosecond_APL2013}.
Thus, the usage of femtosecond laser pulses will give extraordinary freedom in the selection of principal schemes of future hybrid magnonics-photonics circuits.

\section*{ACKNOWLEDGMENTS}
Authors thank Anna V. Kuzikova for the fruitful discussions and technical help during the work process.
P.I.G., Ia.A.F., and N.E.Kh. acknowledge financial support from the Russian Science Foundation (Project 22-22-00326, \url{https://rscf.ru/en/project/22-22-00326/}).


\normalem
\bibliography{apssamp}

\end{document}